# Phenomenological Model of Wetting Charged Dielectric Surfaces and its Testing with Plasma-Treated Polymer Films and Inflatable Balloons

Edward Bormashenko [a,b*], Victor Multanen [b], Gilad Chaniel [a,c], Roman Grynyov[a], Evgeny Shulzinger[a], Roman Pogreb[a], Gene Whyman[a]

[a]*Ariel University, Physics Faculty, 40700, P.O.B. 3, Ariel, Israel*
[b]*Ariel University, Chemical Engineering and Biotechnology Faculty, 40700, P.O.B. 3, Ariel, Israel*
[c]*Bar Ilan University, Physics Faculty, 52900, Ramat Gan, Israel*

E-mail: edward@ariel.ac.il

**Abstract**

Plasma treatment of polymer films results in their electrical charging, which in turn gives rise to an increase in their surface energy. The process results in pronounced hydrophilization of the polymer surfaces. A phenomenological theory relating the change in the apparent contact angle of charged solids to the surface density of the electrical charge is introduced. Partial wetting, inherent for polymer films, becomes possible until the threshold surface density of the electrical charge is gained. The predictions of the theory are illustrated by plasma-treated polymer films and inflatable latex balloons. Deflating the plasma treated latex balloons resulted in an essential increase in the surface charge density of the latex. This increase switched the wetting regime from partial to complete wetting. The kinetics of hydrophobic recovery follows the kinetics of the electrical charge leakage from the surfaces of the plasma treated polymers. The characteristic time of the surface charge leakage coincides with the time scale of the decay of the electret response of plasma treated polymer films.

## 1. Introduction

Wetting of charged surfaces has attracted the attention of researchers in the last decade [1-6]. Electrical charging of surfaces has an effect on contact angles [1, 2, 4], dynamics of the triple line [3, 6] and other interfacial phenomena. The practical interest in the wetting of charged surfaces is based on the fact that they enable, with no mechanical parts, the control of liquid movement and/or a quick change of states of the system [4]. Molecular insights into the hydrophilicity of charged surfaces were discussed recently [1, 5]. The majority of papers devoted to the interfacial properties of charged surfaces deal with conducting surfaces [4, 7-8]. The main (Gibbs) equation relating the surface charge density $\sigma$ to the surface tension $\gamma$, given by $\sigma = -\left(\dfrac{\partial \gamma}{\partial \varphi}\right)_{P,T}$ (where $\varphi, P$ and $T$ are the potential, pressure and temperature correspondingly) was derived for the conducting surfaces [9]. In contrast, our paper is devoted to charged dielectric materials (polymers).

Electrical charging of polymers takes place under plasma treatment (at low and atmospheric-pressure), which is widely used for the modification of surface properties of solid organic materials [10-16]. The plasma treatment creates a complex mixture of surface functionalities which influence surface physical and chemical properties; this results in a dramatic change in the wetting behaviour of the surface [16-22]. Plasma treatment is accompanied by the trapping of plasma ions by a polymer substrate, resulting in its charging. The kinetic model of such charging, which predicts the surface density of an electrical charge supplied by plasma, was reported recently [23]. In the present paper, we propose a phenomenological model, relating the hydrophilization of a plasma-treated polymer surface to its charging. The model was verified with inflatable latex cylindrical balloons enabling the control of the surface charge density.

The pronounced hydrophilization of plasma-treated polymer surfaces is gradually lost with time. This process is called the hydrophobic recovery [22-26]. We relate the hydrophobic recovery to a temporal leakage of the electrical charge.

## 2. Experimental

We used the extruded polypropylene (PP) films with the roughness of the films established with AFM as 10-20 nm; for the AFM study a Park 5 M scanning probe microscope (Scientific Park Instruments) was used and latex-rubber balloons were electrically charged by a plasma unit (EQ-PDC-326 manufactured by MTI Co, USA). Dried compressed air was supplied by Oxygen & Argon Works, Ltd., Israel; moisture was less than 10 ppm, the concentration of oxygen was 20-22%.

Thoroughly cleaned PP film samples, with the dimensions of 25x25 mm and the thickness $h$ of 25 μm, were exposed to an air radiofrequency (RF) plasma discharge under the following parameters: the plasma frequency was 13.56 MHz; the power was 18 W; the pressure was 1 Torr; the volume of the discharge chamber was 840 cm$^3$. The time span of irradiation was 15 s. The latex-rubber balloons were treated under the same parameters. The scheme of the experimental unit used for the plasma treatment of the PP films and latex-rubber balloons is depicted in **Fig. 1**.

The surface charge density of the PP was established with the electrostatic pendulum shown in **Fig. 2a**. The measurement of the angle $\alpha$ between the threads enabled the estimation of the surface charge density $\sigma$. The same method was applied for the estimation of the charge density of the latex balloons charged by plasma, as shown in **Fig. 2b**.

Apparent contact angles were established using the Ramé–Hart goniometer (Model 500). A number of film specimens were simultaneously plasma-treated; this enabled deposition of water droplets on the "freshly treated" portions of the substrates at fixed time intervals. Ten measurements were taken to establish mean apparent contact angles at ambient conditions. The kinetics of a hydrophobic recovery was studied by measuring apparent contact angles every 5 minutes during the first hour, every hour during the first five hours and every day during 3 days. Ten values of the apparent contact angles were taken on both sides of a droplet; the results were averaged.

The study of the influence of the surface charge density was performed with a self-made unit based on inflatable polyisoprene latex cylindrical balloons (Pioneer Balloon Company Qualatex, Canada). The images of the non-inflated and inflated balloons are supplied in **Fig. 3.** The thickness of the balloon wall was 250±50 µm. We measured contact angles on a non-pumped and pumped balloon, and on a balloon deflated after pumping. Apparent contact angles were taken on inflated and deflated balloons before and after a plasma treatment. This procedure allowed control over the surface charge density, which increased after deflation of the balloon.

A ratio of areas was calculated as follows: $A_3/A_2=D_3 \cdot L_3/D_2 \cdot L_2$, where $A_3$, $D_3$, $L_3$ and $A_2$, $D_2$, $L_2$ are the areas, the diameters, and the lengths of pumped and deflated balloons, respectively, as depicted in **Figs. 4-5**. The result was: $A_3/A_2 = 11.4$.

### 3. Results and discussion

### 3.1 Phenomenological model describing wetting of charged dielectric surfaces

Consider the wetting of a polymer surface charged with the constant charge density $\sigma$, as shown in **Fig. 6**, in the situation where the so-called spreading parameter $S$ is negative:

$$S = \gamma_{SA}(\sigma) - (\gamma_{SL}(\sigma) + \gamma) < 0, \tag{1}$$

where $\gamma_{SA}(\sigma)$ and $\gamma_{SL}(\sigma)$ are the surface tensions at the charged solid/air (vapor) and solid/liquid interfaces, respectively, and $\gamma$ is the surface tension at the liquid/air(vapor) interface [8, 27-28]. When $S < 0$, liquid does not spread but forms a cap resting on a substrate with the contact angle $\theta$, as shown in **Fig. 6**. It should be stressed that $\gamma_{SL}(\sigma)$ includes the contribution due to the electrical

charge on the surface. The free energy of a droplet placed on the charged surface is written as [27-29]:

$$G[h(x,y)] = \iint\limits_{S} \left[ \gamma \sqrt{(1+(\nabla h)^2)} + (\gamma_{SL}(\sigma) - \gamma_{SA}(\sigma)) \right] dx dy , \qquad (2)$$

where $h(x,y)$ is the local height of the liquid/air interface above the point $(x,y)$ of the substrate (it is latently supposed that there is no difference between surface tensions and surface energies for $\gamma_{SL}$ and $\gamma_{SA}$), and the integral is extended over the whole substrate area. The first term of the integrand presents the capillary energy of the liquid cap, and the second term describes the change in the energy of the solid substrate covered by liquid. As it was already shown, gravity has no influence on the equilibrium contact angle and, therefore, is neglected in Eq. 2.

We also suppose that the droplet is axisymmetric and does not evaporate; thus, the free energy of the droplet and the condition of the constant volume $V$ are supplied by:

$$G(h,h') = \int\limits_{o}^{a} \left[ 2\pi\gamma x\sqrt{1+h'^2} + 2\pi x(\gamma_{SL}(\sigma) - \gamma_{SA}(\sigma)) \right] dx \qquad (3a)$$

$$V = \int\limits_{o}^{a} 2\pi x h(x) dx = \text{const} . \qquad (3b)$$

where $a$ is the contact radius of the droplet. Exploiting the transversality conditions of the variational problem of wetting, supplied by Eq. 4:

$$(\tilde{G} - h'\tilde{G}'_{h'})_{x=a} = 0 , \qquad (4)$$

where $\tilde{G}(h,h',x) = \gamma x\sqrt{1+h'^2} + x(\gamma_{SL}(\sigma) - \gamma_{SA}(\sigma)) + \lambda x h(x)$, $\lambda$ is the Lagrange multiplier to be deduced from Eq. (3b), yields at the endpoint of the droplet (as explained in detail in Ref. 28-29) the following equation:

$$\cos\theta = \frac{\gamma_{SA}(\sigma) - \gamma_{SL}(\sigma)}{\gamma} . \qquad (5)$$

Expanding $\gamma_{SA}(\sigma)$ and $\gamma_{SL}(\sigma)$ in a Taylor series in powers of $\sigma$, and neglecting terms higher than squared $\sigma$ according to:

$$\gamma_{SA}(\sigma) = \gamma_{SA} + \frac{1}{2}\left(\frac{\partial^2 \gamma_{SA}(\sigma)}{\partial\sigma^2}\right)_{\sigma=0} \sigma^2 + .... \qquad (6a)$$

$$\gamma_{SL}(\sigma) = \gamma_{SL} + \frac{1}{2}\left(\frac{\partial^2 \gamma_{SL}(\sigma)}{\partial\sigma^2}\right)_{\sigma=0} \sigma^2 + ...., \qquad (6b)$$

gives rise to the phenomenological equation:

$$\gamma_{SA}(\sigma) - \gamma_{SL}(\sigma) = \gamma_{SA} - \gamma_{SL} + \frac{\sigma^2}{2\widetilde{C}}, \dots \dots \dots \dots \dots \dots \dots (7)$$

where $\gamma_{SA}$ and $\gamma_{SL}$ are the surface tensions at the non-charged ($\sigma=0$) solid/air (vapor) and solid/liquid interfaces, and $\frac{1}{\widetilde{C}} = \left(\frac{\partial^2 \gamma_{SA}(\sigma)}{\partial \sigma^2}\right)_{\sigma=0} - \left(\frac{\partial^2 \gamma_{SL}(\sigma)}{\partial \sigma^2}\right)_{\sigma=0}$. Indeed, it is reasonable to suggest that the eventual influence of charging on the surface energies will be dependent only even powers of the charge density (the sign of the charge density is not important).

The phenomenological parameter $\widetilde{C}$ has the dimensions of the specific capacity $\frac{F}{m^2}$ (its value will be discussed below in detail). Substituting Eq. 7 into Eq. 5 yields:

$$\cos\theta(\sigma) = \frac{\gamma_{SA} - \gamma_{SL}}{\gamma} + \frac{\sigma^2}{2\gamma\widetilde{C}} = \cos\theta_Y + \frac{\sigma^2}{2\gamma\widetilde{C}}, \qquad (8)$$

where $\theta_Y$ is the equilibrium (Young) contact angle [8, 27-29]. Phenomenological Eq. 8 resembles the well-known Lippmann equation of electrowetting [28-29]. However, the resemblance is only external for several reasons. First of all, the parameter $\widetilde{C}$ by is no means the capacity of the Helmholtz double layer, appearing in the Lippmann equation [7-8]. Secondly, the phenomenological parameter $\widetilde{C}$ reflects the change in the surface energies due to electrical charging of both the solid/air (vapor) and solid/liquid interfaces, as is seen from Eqs. 6a-b. Thus, in spite of its dimensions, it should not be identified with the specific capacity of the double layer. The experimental estimation of its value will be supplied below.

It is seen from Eq. 8 that the condition $\widetilde{C} > 0$ will take place, indeed, the surface charging necessarily decreases the apparent contact angle. This results in the non-trivial thermodynamic inequality $\left(\frac{\partial^2 \gamma_{SA}(\sigma)}{\partial \sigma^2}\right)_{\sigma=0} > \left(\frac{\partial^2 \gamma_{SL}(\sigma)}{\partial \sigma^2}\right)_{\sigma=0}$.

### 3.2. The estimation of the value of the parameter $\widetilde{C}$

When a dielectric surface is charged with the known surface charge density $\sigma$, the value of $\widetilde{C}$ may be established from the measurement of contact angles taken at non-charged and charged surfaces, as is seen from Eq. 8 (of course, within the uncertainty dictated by the contact angle hysteresis [8, 28-33]). We electrically charged PP films with cold air plasma and measured their saturation surface charge densities with an electrostatic pendulum, as explained in detail in the Experimental Section, according to:

$$\sigma_{sat} = \sqrt{2\varepsilon_0 \rho g h \tan\alpha}, \qquad (9)$$

where $\rho$=946 kg/m$^3$ is the density of the PP film, and $h$ is its thickness. Substituting these data, and the experimentally established value of $\tan\alpha$ =0.007-0.013, yields $\sigma_{sat} \cong$ 150-200nC/m$^2$. It was instructive to compare this experimentally established value with that predicted by the kinetic model reported in our recent paper [23]. Our model, based on the assumption that the charging stops when the modulus of the electrical field produced by the charged polymer surface attains the value of the electrical field of the plasma sheath, supplies the saturation value of the surface charge density, estimated as $\sigma_{sat} \cong$ 170nC/m$^2$. It is seen that our model (reported in Ref. 23) accurately estimates the saturation value of the surface charge density [34].

The establishment of the $\sigma$ value allowed estimation of $\tilde{C}$, from the measurement of contact angles before and after plasma treatment, according to Eq. 8:

$$\tilde{C} = \frac{\sigma_{sat}^2}{2\gamma(\cos\theta_{sat}(\sigma_{sat}) - \cos\theta_Y)}. \qquad (10)$$

Substituting $\theta_{sat} = (70 \pm 4)^0; \theta_Y = (102 \pm 2)^0; \gamma = 71\frac{mJ}{m^2}$ into Eq.10 yields: $\tilde{C} \cong (4.0 \pm 0.5)\cdot 10^{-13}$ F/m$^2$. It is recognized that the value of the phenomenological parameter $\tilde{C}$ is much lower than typical values of specific capacities of the Helmholtz layers appearing in the vicinity of metallic electrodes contacting electrolytes [7-8]. Thus, $\tilde{C}$ should not be identified with the specific capacity of the Helmholtz layers.

Consider results obtained with the use of inflatable polyisoprene latex balloons. The equilibrium (Young) contact angle established with the non-treated balloon was $89 \pm 2^0$ (see **Fig. 4a**). It should be stressed that the apparent contact angle remained the same for non-treated pumped and deflated latex balloons; the apparent contact angles before and under the inflation were: $89 \pm 2^0$ and $87 \pm 3^0$ respectively.

Apparent contact angles of inflatable balloons were insensitive to their stretching (inflation) but were sensitive to the plasma treatment. Plasma treatment, performed according to the procedure described in the Experimental Section (see also **Figs. 4a-b**), decreased the apparent contact angle of balloons from $89 \pm 2^0$ before inflation to a value of $72 \pm 2^0$ after inflation. The electrostatic pendulum measurements carried out with the non-inflated balloons supplied the saturation value of the surface charge density $\sigma_{sat}$ =260±40 nC/m$^2$. In this case, the phenomenological parameter $\tilde{C}$, calculated according to Eq. 9, was $\tilde{C} = 1.7\cdot 10^{-12}$ F/m$^2$, and it is one order of magnitude higher that the value of $\tilde{C}$, established for PP. Consider that the phenomenological parameter $\tilde{C}$ comprises the contributions due to the change in charging of both the solid/vapor and solid/liquid interfaces (see Eqs. 6a-b); thus, it may differ essentially for polymers possessing very different chemical structures, such as polypropylene and polyisoprene latex. The plasma treatment induced cleavage of the double carbon-carbon bond in the latex may give rise in the oxidation of the latex, which in turn may influence the value of $\widetilde{C}$ [13].

Afterwards, the plasma-treated balloons were deflated (see **Figs. 4-5**). The specific surface charge density increased significantly, and gained a value of $\sigma \cong \dfrac{A_3}{A_2}\sigma_{sat} \cong 11.4\sigma_{sat} \cong 3\,\dfrac{\mu C}{m^2}$. It is noteworthy, that this surface charge density is still much smaller than the maximal experimentally achievable charge density reported for polymer films in the literature, which is approximately $180\,\dfrac{\mu C}{m^2}$ [35].

Water droplets deposited on the plasma treated deflated balloon demonstrated complete wetting, depicted in **Fig. 4c**, which may reasonably be related to the increase in the surface charge density $\sigma$. Indeed, Eq. 7 works until the value of the surface charge density attains the threshold value $\sigma^*$, given by:

$$\cos\theta_Y + \frac{\sigma^{*2}}{2\gamma\widetilde{C}} = 1. \tag{11}$$

In this case, the spreading parameter $S$ in Eq. (1) attains a zero value. Substituting the above values, and taking into account that in our case $\cos\theta_Y \cong 0$, we obtain $\sigma^* \cong \sqrt{2\gamma\widetilde{C}} \cong 0.5\,\dfrac{\mu C}{m^2}$. It could be recognized that actually $\sigma > \sigma^*$ occurs, i.e. the electrical charge density gained by the deflated latex balloons under the plasma treatment exceeds the threshold value; this in turn gives rise to the complete wetting of plasma-treated deflated balloons. The partial wetting becomes possible when the obvious Eq. 12 takes place (in this case the spreading parameter $S < 0$):

$$\cos\theta_Y + \frac{\sigma^2}{2\gamma\widetilde{C}} < 1. \tag{12}$$

### 3.3. Hydrophobic recovery as the process of leakage of electrical charge

The hydrophilization of plasma-treated polymer surfaces is lost with time. This process is called hydrophobic recovery, and it is accompanied by a diversity of physico-chemical processes, including re-orientation of the hydrophilic moieties of polymer chains [24-26, 36]. We assume that all these processes result in a decrease in the surface charge density, leading in turn to a decrease in the apparent contact angle, as shown in our recent paper [37]. Indeed, the kinetics of the hydrophobic recovery will be phenomenologically described by the following equation (a value of $\widetilde{C} \cong 4 \cdot 10^{-13}$ F/m$^2$ for PP films as it was established in the previous section):

$$\cos\theta(t) = \cos\theta_Y + \frac{\sigma(t)^2}{2\gamma\widetilde{C}}. \tag{13}$$

Thus, the measurement of the kinetics of the apparent contact angle $\theta(t)$ enables the estimation of kinetics of the change in the surface charge density $\sigma(t)$, according to Eq. 14 (resulting from Eq. 13):

$$\sigma(t) = \sqrt{2\gamma\widetilde{C}(\cos\theta(t) - \cos\theta_Y)}. \tag{14}$$

It should be mentioned that the measurement of the kinetics of the surface charge leakage cannot be carried out using the pendulum method, because of its low resolution. Therefore, this measurement was performed using goniometry, according to Eq. (14). The $\sigma(t)$ plot calculated from the study of the kinetics of the hydrophobic recovery of water droplets deposited on plasma treated PP films is supplied in **Fig. 7A**. The experimentally established time dependence of the apparent contact angle $\theta(t)$ is depicted in **Fig. 7B**. The temporal dependence of the surface charge $\sigma(t)$, calculated from Eq. 14, is adjusted by an exponential fit:

$$\sigma(t) = \sigma_\infty + (\sigma_{sat} - \sigma_\infty)e^{-\frac{t}{\tau}}, \tag{15}$$

where $\sigma_\infty$ and $\tau$ are fitting parameters. The fitting parameter $\tau$, representing the characteristic time of a surface-charge leakage, was established roughly as 210 min. This value reasonably coincides with the "long" characteristic time scale of the decay of the electret response of plasma-treated PP films, $\tau_{el} =200$ min, established from electrical measurements and reported recently in Ref. 37. This means that both the electret response and the wetting regime of plasma-treated polymer films are strongly influenced by the presence of the electrical charge on their surfaces.

It should be mentioned that plasma-treated films did not completely restore their initial apparent contact angle. The difference between the initial apparent contact angle and the apparent contact angle measured after one month of storing plasma-treated PP films under ambient conditions was approximately 5º. It is possible to relate this difference to the oxygenation of the polymer surfaces reported by various groups [13, 38] resulting in the residual surface charge density $\sigma_\infty$.

## 4. Conclusions

Wetting of charged surfaces is treated. A phenomenological model relating the equilibrium contact angle to the density of a surface charge deposited on a solid surface is introduced. The model is illustrated by a study of wetting regimes of plasma treated polymer (polypropylene) films and inflatable latex balloons. Plasma treatment electrically charges polymer films and inflated balloons. The surface charge density was established with electrostatic measurements. Plasma treatment of polypropylene films and inflated latex balloons reduced their apparent contact angles. The deflation of latex balloons resulted in the essential increase in the surface charge density. This jump gives rise to the complete wetting of deflated balloons. This observation is described by the introduced phenomenological model predicting a complete-wetting regime for surface charge densities which exceed the threshold value.

We treat the phenomenon of hydrophobic recovery of plasma-treated polymers as the process of leakage of the electrical charge gained by the polymers under plasma treatment. The characteristic time of the change in the surface charge density coincides with the time scale of the decay of the electret response of plasma-treated polymer films.

**Acknowledgements**

**References**

[1]      N. Giovambattista, P. G. Debenedetti, P. J. Rossky, Effect of surface polarity on water contact angle and interfacial hydration structure, J. Phys. Chem. B, 111 (2007) 9581–9587.

[2]      I. S. Bayer, F. Brandi, R. Cingolani, A. Athanassiou, Modification of wetting properties of laser-textured surfaces by depositing triboelectrically charged Teflon particles, Colloid & Polymer Sci. 291 (2013) 367-373.

[3]      A. Niecikowska, M. Krasowska, J. Ralston, K. Malysa, Role of surface charge and hydrophobicity in the three-phase contact formation and wetting film stability under dynamic conditions, J. Phys. Chem. C 116 (2012) 3071−3078.

[4]      L. K. Koopal, Wetting of solid surfaces: fundamentals and charge effects, Adv. Colloid & Interface Sci. 179–182 (2012) 29–42.

[5]      X. Liu, Ch. Leng, L. Yu, K. He, L. J. Brown, Zh. Chen, J. Cho, D. Wang, Ion-Specific Oil repellency of polyelectrolyte multilayers in water: molecular insights into the hydrophilicity of charged surfaces, Angewandte Chemie Intern. Ed. 54 (2015) 4851–4856.

[6]      M. Chakraborty, R. Chatterjee, U. U. Ghosh, S. DasGupta, Electrowetting of partially wetting thin nanofluid films, Langmuir 31 (2015) 4160–4168.

[7]      J. N. Israelachvili, Intermolecular and Surface Forces, Third Edition, Elsevier, Amsterdam, 2011.

[8]      H. Y. Erbi, Surface Chemistry of Solid and Liquid Interfaces, Blackwell, Oxford, 2006.

[9]      L. D. Landau & E. M. Lifshitz Electrodynamics of Continuous Media (V. 8 of A Course of Theoretical Physics), Pergamon Press, Oxford, 1960.

[10]      H. K. Yasuda, J. Wiley & Sons, Plasma polymerization and plasma treatment, New York, 1984.

[11]    M. Strobel, C. S. Lyons, K. L. Mittal (Eds), Plasma surface modification of polymers: relevance to adhesion, VSP, Zeist, The Netherlands, 1994.

[12]    M. Thomas, K. L. Mittal (Eds), Atmospheric Pressure Plasma Treatment of Polymers, Plasma surface modification of polymers: relevance to adhesion, Scrivener Publishing, Wiley, Beverly, USA, 2013.

[13]    M. Lehocky, H. Drnovska, B. Lapcikova, A. M. Barros-Timmons, T. Trindade, M. Zembala, L. Lapcik Jr., Plasma surface modification of polyethylene, Colloids Surf., A 222 (2003) 125-131.

[14]    J. P. Fernández-Blázquez, D. Fell, El. Bonaccurso, A. del Campo, Superhydrophilic and superhydrophobic nanostructured surfaces via plasma treatment, J. Colloid Interface Sci. 357 (2011) 234–238.

[15]    B. Balu, V. Breedveld, D. W. Hess, Fabrication of "roll-off" and "sticky" superhydrophobic cellulose surfaces via plasma processing, Langmuir 24 (2008) 4785–4790.

[16]    Ed. Bormashenko, R. Grynyov, G. Chaniel, H. Taitelbaum, Y. Bormashenko, Robust technique allowing manufacturing superoleophobic surfaces, Appl. Surf. Sci. 270 (2013) 98–103.

[17]    R. M. France, R. D. Short, Plasma treatment of polymers: The Effects of energy transfer from an Argon plasma on the surface chemistry of polystyrene, and polypropylene. A High-Energy resolution X-ray photoelectron spectroscopy study, Langmuir 14 (1998) 4827–4835.

[18]    R. M. France, R. D. Short, Plasma treatment of polymers: Effects of energy transfer from an argon plasma on the surface chemistry of poly(styrene), low density poly(ethylene), poly(propylene) and poly(ethylene terephthalate), J. Chem. Soc., Faraday Trans. 93 (1997) 3173-3178.

[19]    S. Wild, L. L. Kesmodel, High resolution electron energy loss spectroscopy investigation of plasma-modified polystyrene surfaces, J. Vac. Sci. Technol., A 19 (2001) 856-860.

[20]    E. Kondoh, T. Asano, A. Nakashima M. Komatu, Effect of oxygen plasma exposure of porous spin-on-glass films, J. Vac. Sci. Technology. B 18 (2000) 1276-1280.

[21]    D. Hegemann, H. Brunner, Ch. Oehr, Plasma treatment of polymers for surface and adhesion improvement, Nucl. Instr. Meth. Phys. Res., B 208 (2003) 281–286.

[22]    A. Kaminska, H. Kaczmarek, J. Kowalonek, The influence of side groups and polarity of polymers on the kind and effectiveness of their surface modification by air plasma action, Eur. Polym. J. 38 (2002) 1915–1919.

[23]    Bormashenko Ed., Whyman G., Multanen V., Shulzinger E., Chaniel G. Physical mechanisms of interaction of cold plasma with polymer surfaces, J. Colloid & Interface Sci. 448, (2015) 175-179.

[24]    E. Occhiello, M. Morra, F. Garbassi, D. Johnson, P. Humphrey, SSIMS studies of hydrophobic recovery: Oxygen plasma treated PS, Appl. Surf. Sci. 47 (1991) 235-242.

[25]   M. Mortazavi, M. Nosonovsky, A model for diffusion-driven hydrophobic recovery in plasma treated polymers, Appl. Surf. Sci. 258 (2012) 6876–6883.

[26]   E. Bormashenko, G. Chaniel, R. Grynyov, Towards understanding hydrophobic recovery of plasma treated polymers: Storing in high polarity liquids suppresses hydrophobic recovery, Appl. Surf. Sci. 273 (2013) 549– 553.

[27]   de Gennes, P. G.; Brochard-Wyart, F.; Quéré D. Capillarity and Wetting Phenomena, Springer, Berlin, 2003.

[28]   E. Bormashenko, Wetting of Real Surfaces, De Gruyter, Berlin, 2013.

[29]   Bormashenko E. Young, Boruvka-Neumann, Wenzel and Cassie-Baxter equations as the transversality conditions for the variational problem of wetting, Colloids & Surf. A, 345 (2009) 163-165.

[30].   J. F. Joanny, P. G. de Gennes, A model for contact angle hysteresis, J. Chem. Phys. 81 (1984) 552–562.

[31]   C. W. Extrand Y. Kumagai Y., An experimental study of contact angle hysteresis, J. Colloid & Interface Sci. 191 (1997) 378–383.

[32]   R. Tadmor, P. S. Yadav, As-placed contact angles for sessile droplets, J. Colloid & Interface Sci. 317 (2008) 241–246.

[33]   Tadmor R., Approaches in wetting phenomena, Soft Matter (2011) 1577—1580.

[34]   Lieberman, M. A.; Lichtenberg A. J. Principles of plasma discharges and materials processing, J. Wiley & Sons, Hoboken, 2005.

[35]   X. He, H. Guo, X. Yue, J. Gao, Y. Xia, Ch. Hu, Improving energy conversion efficiency for triboelectric nanogenerator with capacitor structure by maximizing surface charge density. Nanoscale 7 (2015) 1896-1903.

[36]   M. Pascual, R. Balart, L. Sanchez, O. Fenollar, O. Calvo, Study of the aging process of corona discharge plasma effects on low density polyethylene film surface, J. Mater. Sci. 43 (2008) 4901–4909.

[37]   Ed. Bormashenko, R. Pogreb, G. Chaniel, V. Multanen, A. Ya. Malkin, Temporal electret behavior of polymer films exposed to cold radiofrequency plasma, Adv. Eng. Mat. 17 (2015) 1175-1179.

[38]   Cl. Riccardi, R. Barni, E. Selli, G. Mazzone, M. R. Massafra, Br. Marcandalli, G. Poletti, Surface modification of poly(ethylene terephthalate) fibers induced by radio frequency air plasma treatment, Appl. Surf. Sci. 211 (2003) 386–397.

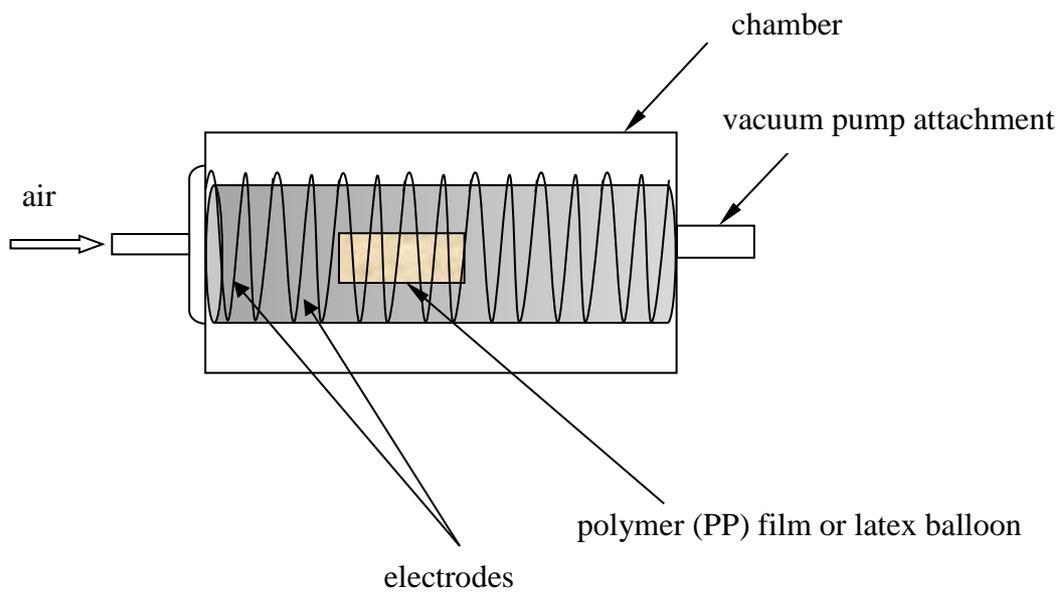

**Fig. 1**. Unit used for plasma charging of polymer films and latex balloons.

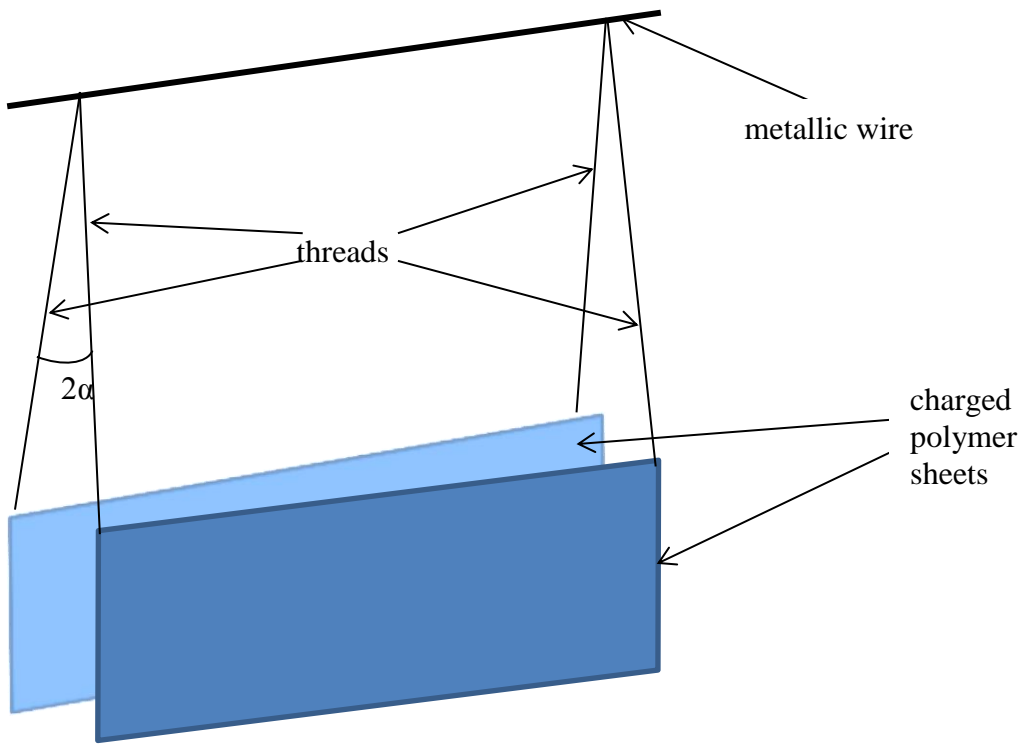

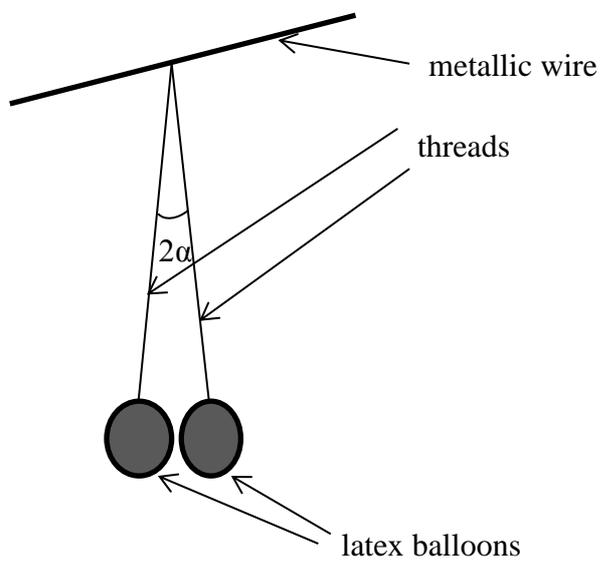

**Fig. 2.** Electrostatic pendulum used for the measurement of the surface charge density of polymers. a) polypropylene (PP) sheets, b) latex rubber balloons.

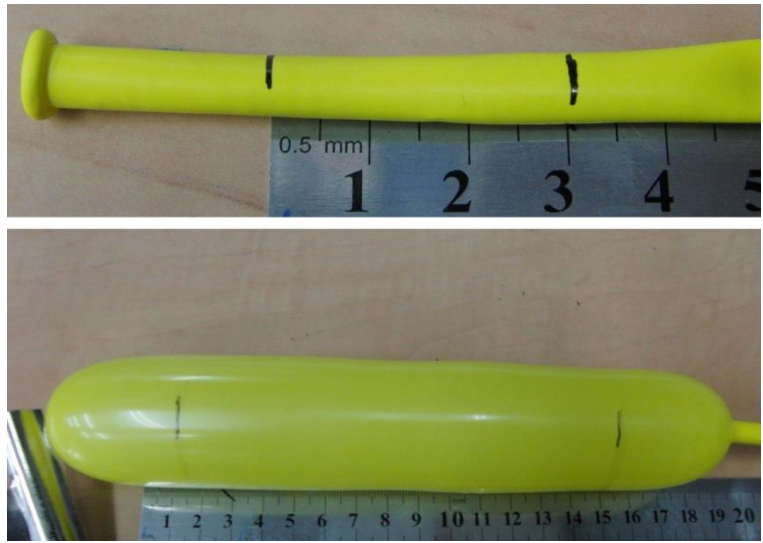

**Fig. 3.** Images of non-inflated (the upper image) and inflated (the lower image) latex balloons.

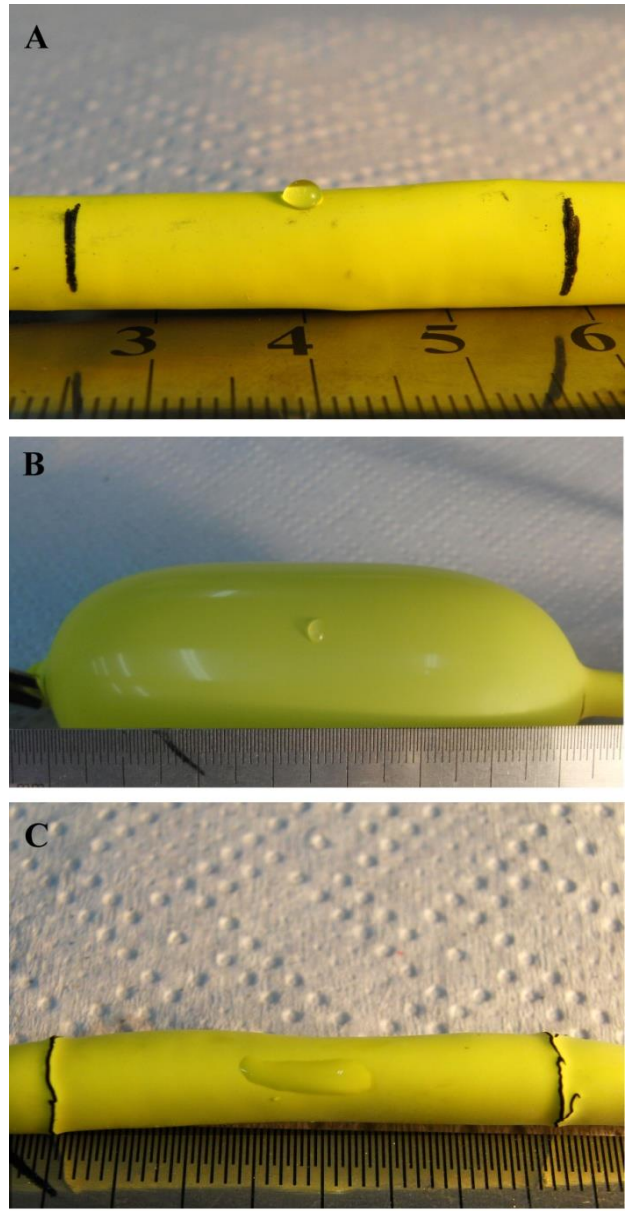

**Fig 4**. Wetting regimes of latex balloons. a) Wetting of a non-treated latex balloon. The apparent contact angle is close to 90º. b) Partial wetting of inflated plasma-treated balloon. The apparent contact angle equals $72^0$. c) Complete wetting of the plasma-treated deflated balloon. The scale of the ruler is 1 cm.

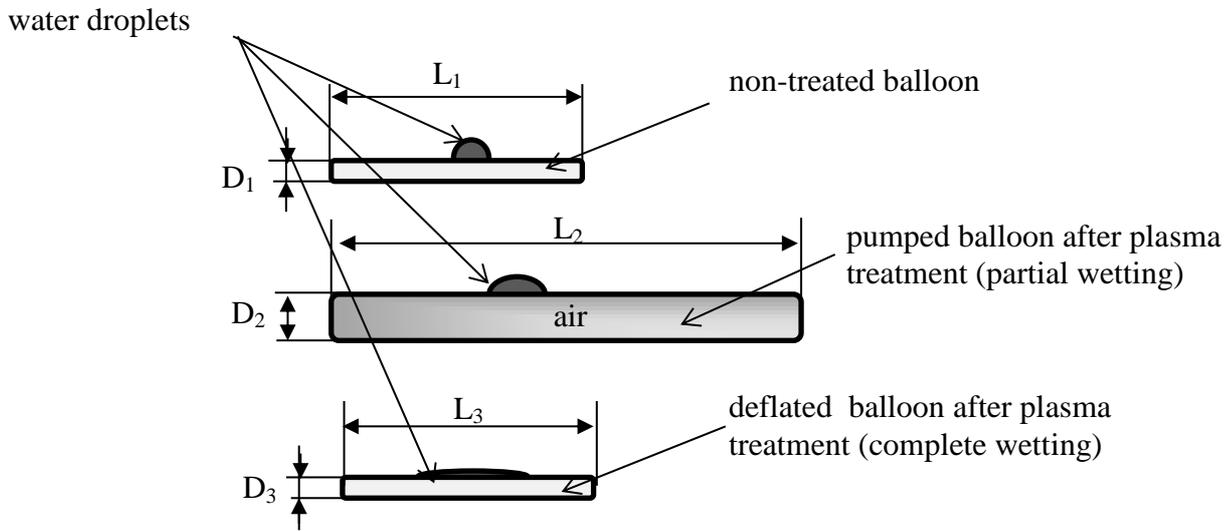

**Fig. 5.** The upper scheme represents a partial wetting of non-inflated non-plasma-treated balloon; the middle scheme shows a partial wetting of inflated plasma-treated balloon; the lower scheme demonstrates a complete wetting of plasma-treated deflated balloon.

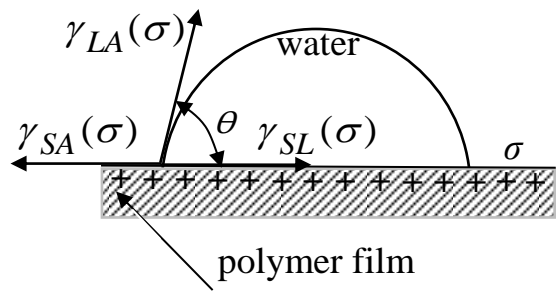

**Fig. 6.** Partial wetting of electrically charged solid with the surface charge density $\sigma$. The equilibrium contact angle $\theta$ is shown.

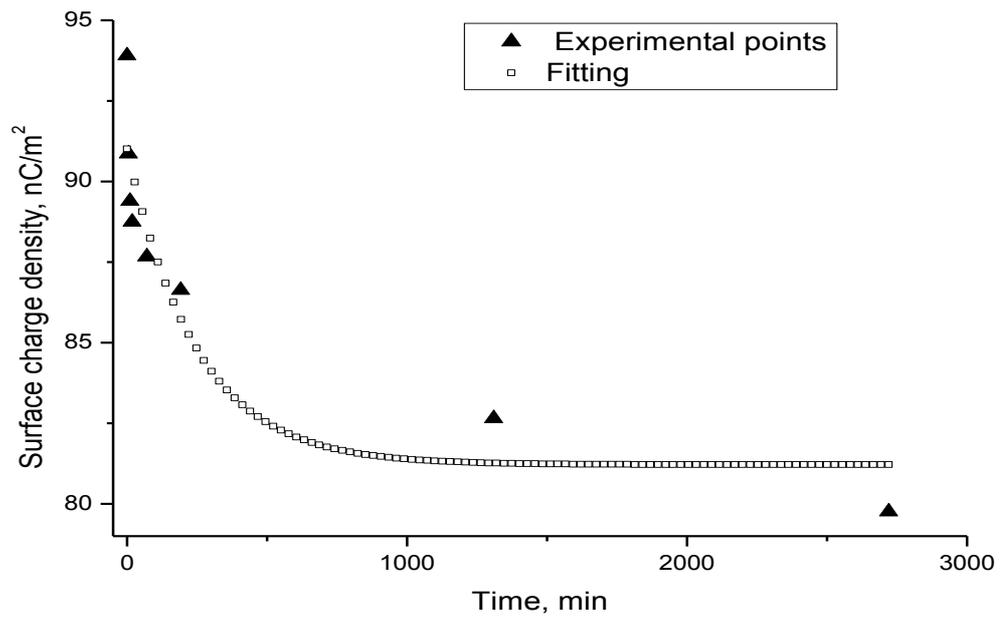

A

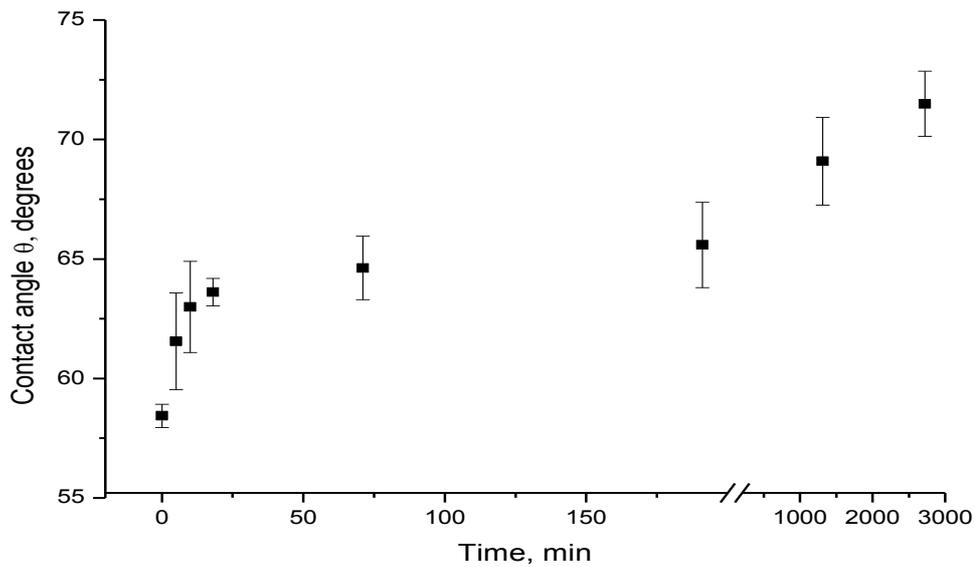

B

**Fig. 7.** Kinetics of hydrophobic recovery of plasma-treated polymer (PP) films.

A) Charge density vs. time. Triangles represent the surface charge density calculated according to Eq. 13. Transparent squares represent the exponential fit performed according to Eq. 14. B) Apparent contact angle $\theta$ vs. time.